# Earlier citation practices are linked to signs of artificial intelligence use in scientific writing

Hongkan Chen[1], Yongyuan He[1], Qingshan Zhou*, and Yi Bu*

Department of Information Management, Peking University, Beijing 100871, China

[1] H.C. and Y.H. contributed equally to this work.

*Corresponding author. Email: zqs@pku.edu.cn or buyi@pku.edu.cn.

**Abstract**

As artificial intelligence (AI) becomes more common in scientific writing, some authors show fewer signs of AI assistance than expected across successive periods. We analysed OpenAlex publication histories for 71,133 authors with at least one eligible first-authored English-language publication each year from 2023 through September 2026. Using abstract-based writing scores, we defined resistance as the predicted minus observed share of abstracts classified as AI-assisted. Positive scores indicate fewer such abstracts than expected. Early resistance was positively associated with later resistance, although the share with positive scores fell. Environmental characteristics contributed more than individual characteristics to predicting later scores, although overall predictive power was low. Having a collaborator with a non-positive early score was associated with lower subsequent resistance. Authors with positive mean scores in both 2023–2024 and 2025–2026 cited older work, drew on fewer subfields and more often cited sources they had used before than authors with non-positive mean scores in both periods. These differences were evident before 2023. The adjusted gap in reference reuse rose from 3.24 percentage points before 2023 to 7.95 in 2025–2026. The findings link variation in signs of AI assistance to earlier differences in how scientists used prior work.

## Introduction

Generative artificial intelligence (AI) can assist with tasks ranging from programming and data analysis to manuscript preparation (1, 2). Making a tool available, however, does not tell us how readily it will become part of a researcher's routine. Diffusion of innovations theory links adoption to expected benefits, compatibility with existing practices and what people learn from others' experiences (3). Research on innovation resistance further emphasises that people may decline or defer a technology when it disrupts familiar routines or conflicts with their beliefs (4, 5). Academic writing brings the question of fit into focus. As they write, researchers organise evidence and develop the arguments through which they communicate their results (6). In an experiment on professional writing, AI reduced the time people spent on writing tasks (7). In science, the prospect of saving time raises a question about which parts of writing authors are willing to delegate.

Empirical studies have identified differences in estimated uptake of AI-assisted academic writing. Estimates vary across countries (8, 9). In biomedical research, estimated use grew faster among earlier-career authors and in countries with lower English proficiency (10). Estimated use also varies across disciplines (11, 12) and is associated with smaller teams and lower mean academic ages (13). A survey of researchers found that acceptance of generative AI varies across research tasks, with language editing generally viewed more positively than some other uses (14). In a survey of editors and editorial board members in history, political science and economics, about a third of respondents considered it unacceptable for authors to use AI to turn notes into paragraphs, with responses varying by discipline (15). Some respondents emphasised that writing itself contributes to developing scholarly ideas and arguments (16). This evidence tells us where AI-associated writing is more common and which uses people find acceptable. When authors initially show less evidence of AI assistance than expected, how often does that pattern persist as the technology spreads?

Research on technology acceptance links reluctance to adopt a new system to habits formed around an existing one (17). In academic writing, established routines include how researchers decide which sources to consult, which claims need further evidence and which studies to return to. Disciplinary conventions and collaborations may also shape these decisions. Recent work links AI-associated writing to the age and variety of cited work (18). A preprint reports that authors who later showed stronger AI-writing signals already had more interdisciplinary publication portfolios (19). We ask whether relatively low levels of AI-associated writing persist and whether persistence is associated with how authors drew on prior work. Publications from before 2023 allow us to compare authors' earlier citation practices, including how often they returned to specific sources they had previously cited.

We follow 71,133 authors from 2023 to 2026, combining publication records from OpenAlex (20) with abstract-based writing scores for their first-authored publications. We call the predicted minus observed share of abstracts classified as AI-assisted the resistance score. The prediction accounts for individual, environmental and publication characteristics. A positive score means fewer abstracts were classified as AI-assisted than predicted. A non-positive score means the observed share met or exceeded the prediction. The score describes an abstract-level signal, while authors' intentions and tool use remain unobserved. We also assess persistence using an absolute threshold for authors' mean AI-associated writing scores (see Materials and Methods). We examine the prevalence and persistence of resistance and how subsequent resistance relates to authors' individual characteristics, research environments and collaborators' resistance scores. For the reference-use comparison, we contrast authors with positive mean scores in both 2023–2024 and 2025–2026 with authors whose mean scores were non-positive in both periods. Differences between these groups were already visible in publications before 2023. Authors who later scored positively in both periods had cited older work, drawn on fewer subfields and returned more often to sources they had cited before.

## Results

### Changes in resistance to AI-assisted writing

Resistance in 2023–2024 was positively associated with resistance in 2025–2026 (Fig. 1a). At each transition, more authors moved out of the resistant group than into it. The share of the full cohort with a positive score fell from 52.9% in 2023–2024 to 44.1% in 2025 and 37.2% in 2026 (Fig. 1b). Using an absolute AI-associated writing score threshold yielded different proportions but likewise showed a decline in the share of authors below the threshold (SI Appendix, Fig. S8b). Initial prevalence and continuous resistance varied across disciplines (SI Appendix, Fig. S1). Over half of initially resistant authors in mathematics and in the combined physics and astronomy field also had positive scores in both 2025 and 2026. The corresponding shares were about a quarter or less in computer science, engineering and business. The proportion of authors remaining continuously resistant was higher in the English-speaking institutional country category than in the non-English-speaking category (SI Appendix, Fig. S2a). Among initially resistant authors, subsequent resistance was also lower in the non-English-speaking category in both years (SI Appendix, Fig. S2b).

## Individual and environmental correlates of resistance

We next examined how individual and environmental characteristics predicted authors' subsequent resistance. Among initially resistant authors, environmental characteristics contributed more to out-of-sample prediction of subsequent resistance than individual characteristics. Models using early resistance alone explained 2.4% of the variation in resistance in 2025 and 1.4% in 2026. Adding individual and environmental characteristics increased these values to 4.0% and 2.7%, respectively. Environmental characteristics contributed approximately twice as much as individual characteristics to the gain in out-of-sample predictive performance in 2025 and more than three times as much in 2026 (Fig. 1c; SI Appendix, Table S1). Among initially resistant authors with at least one measurable collaborator, those who had a collaborator with a non-positive early score showed lower subsequent resistance than those whose measurable collaborators all had positive early scores (Fig. 1d; SI Appendix, Table S2). Their resistance scores were lower by an estimated 0.060 in 2025 and 0.052 in 2026. The association was also statistically significant when only collaborators who had served as corresponding authors were considered.

## Reference use by early and late resistance status

We compared reference use between authors with positive mean resistance in both periods and those with non-positive mean resistance in both. The former group had higher mean and median reference ages in each of the three publication windows (Fig. 2a; SI Appendix, Table S3). Adjusted mean reference age changed little in the former group, from 10.50 years before 2023 to 10.60 in 2025–2026. In the comparison group, it declined from 9.75 to 9.01 years, with most of the decrease evident by 2023–2024. The age distribution of references also differed between the groups (SI Appendix, Table S3). In 2025–2026, references aged 0–7 years accounted for a smaller share in the group with positive scores in both periods. References aged 8–19 years and at least 20 years accounted for larger shares. Reference variety and diversity were lower in this group in each publication window (Fig. 2b and c). By 2025–2026, the group differences reached 0.71 subfields and 0.030 on the diversity index, respectively, more than twice the corresponding gaps before 2023. Reference balance was also lower in this group in 2023–2024 and 2025–2026 (SI Appendix, Table S3). Overall, these authors drew on a narrower range of disciplinary sources, with larger differences between the groups after 2023.

Authors positive in both periods also cited a higher share of references they had used in earlier first-authored publications (Fig. 2d and SI Appendix, Table S4). The adjusted gap rose from 3.24 percentage points before 2023 to 7.95 in 2025–2026. In 2023–2024 and 2025–2026, they also cited larger shares of references from two fixed libraries, one covering five calendar years from each author's first observed first-authored publication year and the other covering all their first-authored publications before 2023 (SI Appendix, Table S4). In the three-group comparison, authors whose score changed from positive in the early period to non-positive in the late period had lower reference reuse than those positive in both. This gap was already evident before 2023 and was larger in 2025–2026 (SI Appendix, Fig. S3f).

## Discussion

The differences in citation practices preceded the widespread availability of generative AI. Authors who went on to record positive resistance scores in both periods had already cited older literature, drawn on fewer subfields and returned more often to their previously cited sources. Resistance in 2023–2024 was associated with subsequent resistance even as the share of authors classified as resistant fell. Diffusion of innovations theory treats compatibility with existing practices as one condition for uptake (3). Innovation resistance research distinguishes reluctance rooted in established routines from objections to the technology itself (21). In scientific writing, the distinction matters because researchers can use AI to edit prose while retaining control over how they weigh evidence and develop an argument. Our data trace relative writing patterns over time. Identifying which parts of writing authors delegate, or why they choose to do so, would require evidence from the writing process itself.

Among initially resistant authors, environmental characteristics contributed more to out-of-sample prediction than individual characteristics after accounting for early resistance. Even with both sets of characteristics, the models accounted for 4.0% of the variation in 2025 and 2.7% in 2026. The association between having a measurable collaborator with a non-positive early score and lower subsequent resistance parallels earlier work linking collaborators' AI experience to the use of AI in research (22). Shared drafting and revision could introduce authors to AI tools. A collaborator could also change the language of a joint abstract while leaving the focal author's own writing practices unchanged. A further possibility is that authors select collaborators who work in similar ways (23). Disciplinary differences may reflect variation in research tasks, writing conventions and judgements about which forms of AI assistance are acceptable (14, 24). Privacy and security may also constrain use when manuscripts contain sensitive or unpublished material (25). Draft histories and records of how collaborations formed could help distinguish peer learning, shared drafting and collaborator selection.

Reference age describes how far back authors draw on the literature. Reference reuse adds a measure of continuity within an author's own publication history. The direction of the reference-age difference accords with previous evidence that authors identified as adopters of large language models cite more recent work (18). The age gap widened mainly because the references of authors with non-positive scores in both periods became more recent. The three-group comparison also distinguishes authors who changed status. Before 2023, authors who later switched from positive to non-positive scores had already reused a smaller share of references than authors who scored positively in both periods. The two-group comparisons show where the reuse gap widened. The adjusted gap for sources in the pre-2023 library was similar in the two later periods. For sources previously cited outside that library, it rose from 0.74 to 2.35 percentage points (SI Appendix, Table S4). The group difference in reuse thus extended to sources first cited in the authors' observed publication histories from 2023 onwards. The risks and rewards of scientific work can encourage authors to pursue established lines of inquiry (26), while reliance on older literature tends to increase as scientific careers progress (27). Returning to familiar work may help authors pursue unresolved questions across successive studies. These citation patterns can also arise when authors remain in the same research area or when fields differ in how quickly their literatures turn over.

An AI writing tool could start with the sources a researcher already uses, link each suggested comparison to the underlying texts and help the researcher find relevant evidence outside that collection. This would let authors check whether a suggestion represents earlier work accurately and whether it overlooks a competing explanation. Evaluating such a tool would require testing how well authors select and check sources as well as measuring any time saved in drafting.

The differences observed before 2023 have implications for how changes in scientific citation are attributed to AI assistance. If authors with different citation histories take up these tools at different rates, the mix of references in AI-assisted papers can shift even when individual authors' citation practices stay the same. A study could use authors' publication histories to establish their starting points and randomize access to AI assistance for comparable writing tasks. Recording which sources authors consult, which they cite and how they use them in a draft would help distinguish changes in literature search from changes in the evidence used to support a claim. It would also allow researchers to test whether the same form of assistance has different effects for authors who often return to sources they have cited before.

AI-associated writing scores measure resemblance to reference texts in published abstracts. Disciplinary language conventions, editing practices and changes in generative models may all affect that resemblance. The resistance measure also depends on the classification threshold (28, 29) and the model used to estimate expected probabilities. Using an absolute score threshold checks sensitivity to the choice of resistance measure while retaining the same underlying text signal. Surveys, interviews or drafting records could help distinguish opposition, restricted access and satisfaction with existing methods. The pre-2023 comparisons are retrospective and use groups defined by the signs of their subsequent period scores. They establish that the differences in citation practices were already present. Whether those practices affect later uptake remains an open question. The cohort includes only authors with at least one eligible first-authored English-language publication in every year from 2023 to 2026. The observation window for 2026 ends on 23 September. The results may differ for authors entering science, publishing intermittently or publishing mainly in other languages.

## Materials and Methods

**Data collection and processing.** We constructed an author cohort using the 23 September 2026 snapshot of OpenAlex by retaining authors with at least one first-authored English-language article, review or conference paper in each year from 2023 to 2026. Authors with more than 30 eligible first-authored publications during the observation period were excluded, yielding a final cohort of 71,133 authors and 565,602 eligible publications. Every author had at least one publication with a scorable abstract in each observation year. Historical measures were derived from all available prior first-authored publications for each author. We estimated AI-associated writing scores, $\alpha$, for publication abstracts using maximum likelihood estimation (18, 30). Scores range from zero to one, with higher values indicating greater similarity to AI reference texts (SI Appendix, Note 1). We generated the AI reference texts using GPT-3.5 Turbo and used Qwen2.5-72B-Instruct in a sensitivity analysis. During manuscript preparation, the authors used GPT-6.1 for selected coding tasks and reviewed and verified the outputs.

**Resistance measures.** We defined the binary indicator of AI-associated writing for each publication $p$ as $Y_p = 1\ (\alpha_p \geq 0.1)$. Using publications from 2023–2024, we fitted a ridge-penalised logistic model:

$$logit\ Pr\left(Y_p = 1 \mid \boldsymbol{X}_p\right) = \beta_0 + \boldsymbol{\beta}^T \boldsymbol{X}_p \quad (1)$$

The predictor vector $X_{\boldsymbol{p}}$ included field, the institutional country category, publication date, academic age, cumulative first-author output and output in the preceding 12 months, team size, and mean $\alpha$ before 2023. Authors were divided into three groups, and each group's publications were predicted using models fitted to the other two groups. For 2025 and 2026, predictions were calibrated using same-year publications from the other two author groups to account for changes over time in the share of abstracts classified as AI-assisted. For each author and year, resistance was calculated as the average difference between predicted probabilities and observed classifications:

$$R_{it} = \frac{1}{n_{it}} \sum_{p \in \mathcal{P}_{it}} \left(\hat{p}_p^{\text{cal}} - Y_p\right) \quad (2)$$

The average was taken over author $i$ 's $n_{it}$ first-authored publications in year $t$. The predicted probability $\hat{p}_p^{\text{cal}}$ incorporates calibration for 2025–2026. Positive resistance indicates a lower proportion of abstracts classified as AI-assisted than predicted. We calculated early resistance for 2023–2024 and late resistance for 2025–2026 by giving the two annual scores in each period equal weight. Continuous resistance through 2025 required a positive early-period score and a positive 2025 score. Continuous resistance through 2026 also required a positive 2026 score. Separately, we tracked transitions between groups defined by whether mean $\alpha$ was below 0.10 (18) (SI Appendix, Section S2.4). We applied the threshold to the 2023–2024 mean and then to the 2025 and 2026 annual means separately.

**Predicting subsequent resistance.** Among authors with early resistance, we fitted separate models for resistance in 2025 and 2026. Individual characteristics included academic age, cumulative and recent first-author output, publication count during 2023–2024, and mean $\alpha$ before 2023. Environmental characteristics included field, the institutional country category, distinct collaborator count and mean team size. We compared four models: a baseline containing a spline of early resistance, the baseline plus individual characteristics, the baseline plus environmental characteristics, and a full model containing both groups. Grouped Shapley decomposition divided the improvement in predictive $R^2$ beyond the baseline between the two groups of characteristics. These comparisons assessed whether individual characteristics or research environments contributed more to predicting subsequent resistance out of sample (SI Appendix, Section S3.1).

**Collaborator associations.** Among initially resistant authors with at least one measurable collaborator in 2023–2024, we compared authors with and without a collaborator whose early resistance score was non-positive. We calculated collaborators' scores from their own first-authored publications, excluding papers shared with the focal author. A collaborator needed at least three remaining publications and a calculable early score. We fitted separate annual regressions for all collaborators and for collaborators who had served as corresponding authors (SI Appendix, Section S3.2).

**Patterns of reference use.** We compared authors with positive mean resistance in both 2023–2024 and 2025–2026 with authors with non-positive mean resistance in both periods. Outcomes covered the age, disciplinary composition and reuse of cited literature. For each outcome, we fitted separate author-level ordinary least squares models for publications before 2023, in 2023–2024 and in 2025–2026. In supplementary analyses, we also included authors with positive early and non-positive late scores (SI Appendix, Note 4).

## Acknowledgments

During manuscript preparation, the authors used GPT-6.1 to improve the language of portions of the manuscript. The authors reviewed and verified all AI-assisted content and take full responsibility for the final manuscript. We thank all members of the Knowledge Discovery Lab at Peking University for their support with the analyses and constructive comments. This work was supported by the National Natural Science Foundation of China (#72474009).

## Data and code availability

The data used in this study were obtained from OpenAlex (20) (https://openalex.org). The code used for data processing and analysis is available at https://github.com/pku-hyy/ai_writing_resistance.

## Figures

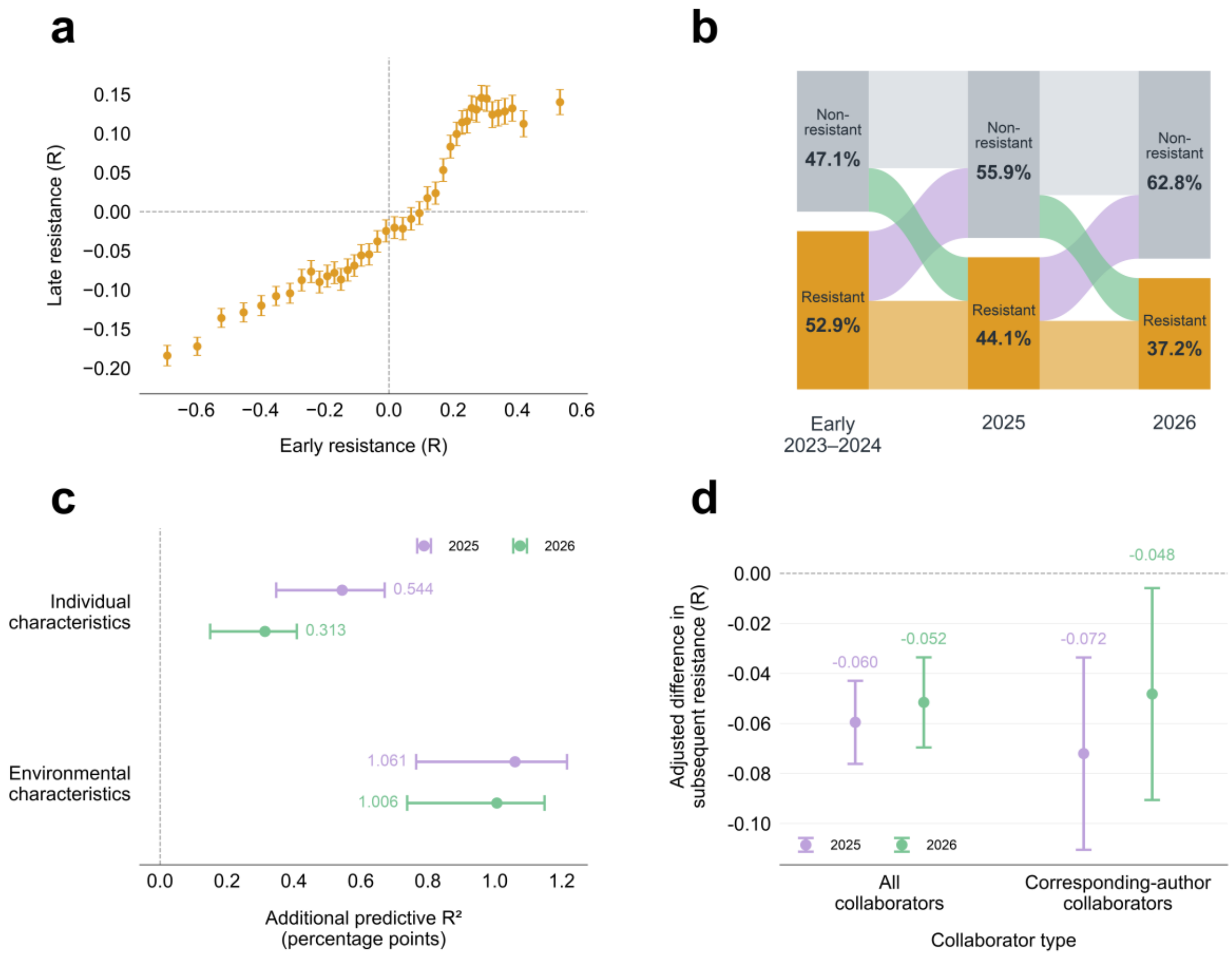

**Fig. 1. Changes in resistance to AI-assisted writing and its correlates.** **a**, Binned association between early (2023–2024) and late (2025–2026) resistance. **b**, Transitions between resistance states across the early period (2023–2024), 2025 and 2026. Percentages refer to the full cohort at each time point. Positive scores determine resistance status at that time point. Purple and green flows indicate exits from and entries into the resistant group, respectively. **c**, Additional predictive contributions of individual and environmental characteristics. **d**, Adjusted associations between having a measurable collaborator with a non-positive early score and subsequent resistance, by collaborator type. Error bars indicate 95% confidence intervals.

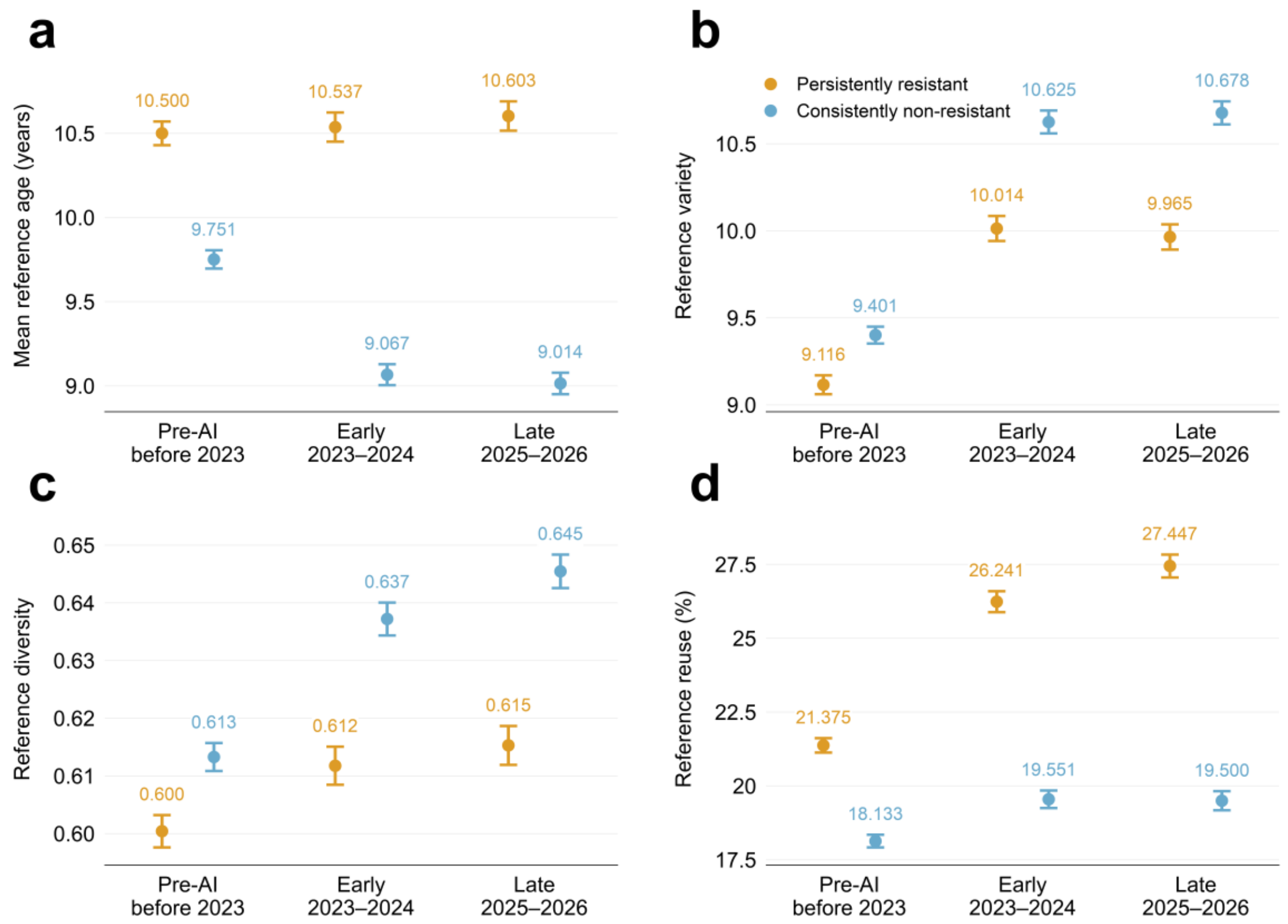

**Fig. 2. Reference use by early and late resistance status. a**, Mean reference age. **b**, Reference variety. **c**, Reference diversity. **d**, Reference reuse. The orange group (labelled "Persistently resistant") had positive mean scores in both periods. The blue group (labelled "Consistently non-resistant") had non-positive mean scores in both periods. "Pre-AI" denotes publications before 2023. Points show adjusted means. Error bars indicate 95% confidence intervals.

Supporting Information for

# Earlier citation practices are linked to signs of artificial intelligence use in scientific writing

Hongkan Chen[1], Yongyuan He[1], Qingshan Zhou*, and Yi Bu*

Department of Information Management, Peking University, Beijing 100871, China

[1] H.C. and Y.H. contributed equally to this work.

*Corresponding author. Email: zqs@pku.edu.cn or buyi@pku.edu.cn.

## Supplementary Note 1: Estimation of AI-associated writing scores

### S1.1 Construction of human and AI reference texts

We sampled 7,200 abstracts published in 2020–2021 from six fields: medicine, engineering, social sciences, computer science, physics and astronomy, and arts and humanities. We randomly selected 50 abstracts containing at least 100 words in each field–month combination.

We generated an AI counterpart for each abstract in two separate calls to GPT-3.5 Turbo. The first call extracted content notes using the following instruction:

> "Convert the following scientific abstract into concise, ordered bullet-point notes. Preserve its substantive information, numerical values, technical terms and levels of certainty. Minimize copying of the original wording and do not add information."

The second call received only the notes and the original abstract's word count. The instruction was:

> "Write a coherent, single-paragraph scientific abstract using only the following notes. Preserve all substantive information, numerical values, technical terms and levels of certainty. Do not add information. Keep the length within 70–130% of the original abstract's word count. Return only the abstract."

### S1.2 Estimation of abstract-level scores

We estimated the abstract-level score $\alpha$ from fixed reference distributions derived from 62,971 human sentences and 40,785 AI sentences. The vocabulary contained 14,655 words occurring in at least five human sentences and three AI sentences. For word $w$, the probabilities of sentence-level occurrence were:

$$P_w = \frac{c_{H,w} + 0.5}{N_H + 1}, \qquad Q_w = \frac{c_{A,w} + 0.5}{N_A + 1} \tag{1}$$

where $c_{H,w}$ and $c_{A,w}$ denote the numbers of reference sentences containing the word, and $N_H$ and $N_A$ are the corresponding numbers of sentences. Repeated occurrences within a sentence were counted once.

Abstracts were cleaned of HTML markup and excess whitespace, segmented into sentences, lowercased and tokenized. Tokens consisting only of numbers were excluded. Let $x_{sw}$ indicate the presence of word $w$ in sentence $s$. Let $V$ denote the selected vocabulary. Under the model, word occurrences were conditionally independent within a sentence. The reference likelihoods were:

$$P_s = \prod_{w \in V} P_w^{x_{sw}} (1 - P_w)^{1 - x_{sw}}, \qquad Q_s = \prod_{w \in V} Q_w^{x_{sw}} (1 - Q_w)^{1 - x_{sw}} \tag{2}$$

For publication $p$, we estimated:

$$\hat{\alpha}_p = \underset{\alpha \in [0,1]}{\operatorname{argmax}} \sum_{s \in p} \log\big((1 - \alpha) P_s + \alpha Q_s\big) \tag{3}$$

The score is the estimated weight of the AI-reference component of the mixture.

### S1.3 Robustness to the model used to generate AI reference texts

We generated AI reference texts with Qwen2.5-72B-Instruct using the same human abstracts and procedure as for GPT-3.5 and recalculated the AI-associated writing scores. The two sets of scores were positively correlated (Pearson's $r = 0.750$; Spearman's $\rho = 0.772$). Mean $\alpha$ was 0.453 with the Qwen reference texts and 0.238 with the GPT-3.5 reference texts (Figure S6).

## Supplementary Note 2: Construction of resistance measures

### S2.1 Record processing and variable definitions

The variables used in this study were defined as follows.

1) **Field** was obtained from the field associated with each publication's primary topic in OpenAlex, with an author's field defined as the most frequent field among the author's 2023–2024 publications. We compared initial prevalence and continuous resistance across fields (Figure S1). Initial prevalence was the share of all authors in each field with positive early resistance. Continuous resistance through 2025 or 2026 was the share of initially resistant authors with a positive score in each subsequent year through the year indicated.

2) **The institutional country category** was the share of the focal author's distinct affiliation countries listed on a publication that were classified as English-speaking, with each country receiving equal weight. English-speaking countries comprised Australia, Canada, Ireland, New Zealand, the United Kingdom and the United States.

3) **Publication date** was represented by a calendar-quarter index.

4) **Academic age** was calculated as the publication year minus the year of the author's first observed first-authored publication.

5) **Cumulative output** counted all preceding first-authored publications, while **recent output** counted those published during the preceding 12 months.

6) **Team size** was the number of authors listed on a publication, and **collaborator count** was the number of distinct coauthors appearing on the author's first-authored publications during 2023–2024.

7) **Historical mean $\boldsymbol{\alpha}$** was the average AI-associated writing score of an author's scorable first-authored publications before 2023. We included this measure because some authors' abstracts already exhibited features of the AI reference texts before these tools became widely available.

8) **Early publication count** was the number of first-authored publications during 2023–2024.

### S2.2 Early-period prediction

Authors were assigned to three groups by hashing their OpenAlex IDs with SHA-256 and a fixed seed (20261018), converting the first 12 hexadecimal digits to an integer and taking the remainder on division by three. The groups contained 23,716, 23,449 and 23,968 authors. All publications by the same author remained in the same group across years. For each target group, we fitted a ridge-penalised logistic model using the 2023–2024 publications from the other two groups and generated predicted probabilities for the target group's publications. Training weights gave each author equal total weight across their publications. The predictor vector included field, the institutional country category, publication date, academic age, cumulative and recent output, team size, and historical mean $\alpha$. Output counts and team size were transformed using $\log(1+x)$. Academic age and transformed output counts entered as linear and quadratic terms. We standardised numerical predictors using the training data. The predicted probability was:

$$\hat{p}_p^{(0)} = \text{logit}^{-1}\left(\hat{\beta}_0 + \boldsymbol{X}_{\boldsymbol{p}}^{\boldsymbol{T}}\hat{\boldsymbol{\beta}}\right) \tag{4}$$

The fitted early-period models were retained for predictions in 2025 and 2026.

### S2.3 Calibration for 2025–2026

Calibration used publications from the same year in the other two author groups. Initial predicted probabilities for these publications were obtained from the same early-period model used for the target group.

Prediction errors were calculated as $e_p = Y_p - \hat{p}_p^{(0)}$. Field–quarter means, $\bar{e}_{fq}$, and quarter means, $\bar{e}_q$, were calculated by first averaging errors within each author and then averaging across authors, giving each author equal weight. Field–quarter means were then shrunk towards their corresponding quarter means:

$$\hat{\delta}_{fq} = \frac{n_{fq}\bar{e}_{fq} + 20\bar{e}_q}{n_{fq} + 20}, \tag{5}$$

where $n_{fq}$ is the number of contributing authors in field $f$ and quarter $q$.

Second, we fitted a ridge regression to the residual prediction errors after the first adjustment. Predictors included field–quarter categories, the institutional country category, a spline of academic age, and categories of cumulative and recent output, team size, the probability after the first adjustment and historical mean $\alpha$. Authors received equal total weight, and the ridge penalty was 100.

Let $\hat{g}_p$ denote the predicted error from this regression. The adjusted probabilities were:

$$\begin{aligned} \hat{p}_p^{(1)} &= clip_{[0,1]}\left(\hat{p}_p^{(0)} + \hat{\delta}_{f(p),q(p)}\right), \\ \hat{p}_p^{cal} &= clip_{[0,1]}\left(\hat{p}_p^{(1)} + \hat{g}_p\right). \end{aligned} \tag{6}$$

The second model's outcome was $Y_p - \hat{p}_p^{(1)}$. Probabilities were restricted to $[0,1]$ after each adjustment. For 2023–2024, $\hat{p}_p^{cal}$ denotes the original prediction from the model fitted to the other author groups. We calculated annual resistance by averaging predicted probabilities minus observed binary classifications within each author-year. We gave the two annual scores equal weight in the early (2023–2024) and late (2025–2026) periods.

### S2.4 Robustness using absolute writing scores

We classified an author as having a low absolute score when their mean $\alpha < 0.10$, and assigned authors with mean $\alpha \geq 0.10$ to the comparison category. Early- and late-period scores were calculated by averaging annual means equally within each period. For the transition analysis, we applied the threshold to the 2023–2024 mean and separately to the 2025 and 2026 annual means. We repeated the persistence analyses, the comparison of individual and environmental contributions, and the collaborator analyses using the absolute score criterion. We applied the corresponding eligibility rules and used the same control variables (Figure S8). The prediction models using the absolute score used spline knots at 0.025 and 0.06 for the early mean score.

### S2.5 Persistence across the publication sequence

Because authors published different numbers of papers each year, we also examined persistence using equally sized blocks of publications for each author. We selected 3,777 authors with 15–30 eligible first-authored publications during 2023–2026 and at least five in each of the early and late periods. We ordered publications by date and compared mean resistance in each author's first and last five publications. We excluded authors if a block boundary split publications with the same date. Resistance in the first five publications was positively associated with resistance in the last five (Spearman's $\rho = 0.299$; Figure S7a). Among authors with positive resistance in the first five publications, 53.1% also had positive resistance in the last five, compared with 29.5% among those whose initial resistance was non-positive (Figure S7b).

## Supplementary Note 3: Predictive models and collaborator associations

### S3.1 Predictive model specifications

For year $t$, we modelled subsequent resistance among authors with positive early resistance as follows:

$$R_{it} = \beta_{0t} + f_t(R_{i,E}) + \boldsymbol{A}_i^T \boldsymbol{\beta}_{A,t} + \boldsymbol{E}_i^T \boldsymbol{\beta}_{E,t} + \varepsilon_{it} \quad (7)$$

where $\boldsymbol{A}_i$ and $\mathbf{E}_i$ contain the individual and environmental predictors. $R_{i,E}$ denotes the author's early resistance. Individual characteristics included academic age, cumulative and recent output, publication count during 2023–2024, and historical mean $\alpha$. Environmental characteristics included field, the institutional country category, distinct collaborator count and mean team size. The early-resistance term $f_t$ was an unpenalised cubic spline with interior knots at 0.1 and 0.3. Models with individual predictors included academic age as a cubic spline with four degrees of freedom.

We compared the baseline model containing the early-resistance spline ($M_0$), the baseline plus individual characteristics ($M_A$), the baseline plus environmental characteristics ($M_E$), and the full model ($M_{AE}$). All four models used identical author groups. For each model with additional predictors, we used cross-validation within the two training groups to select a ridge penalty from 1, 10, 100 and 1,000. We refitted each model on both training groups and evaluated it on the third.

For model $m$, we calculated $R^2$ by comparing each author's observed resistance with predictions from models trained on the other author groups:

$$R_m^2 = 1 - \frac{\sum_i (R_{it} - \hat{R}_{it,m})^2}{\sum_i (R_{it} - \bar{R}_t)^2} \quad (8)$$

where $\bar{R}_t$ is the mean observed resistance in the analysis sample for year $t$.

The two-group Shapley contributions were:

$$\phi_A = \frac{1}{2}[(R_A^2 - R_0^2) + (R_{AE}^2 - R_E^2)] \quad (9)$$

$$\phi_E = \frac{1}{2}[(R_E^2 - R_0^2) + (R_{AE}^2 - R_A^2)] \quad (10)$$

Their sum equals the full model's improvement over the baseline. Contributions were expressed in $R^2$ percentage points. We constructed 95% confidence intervals from 200 author-level bootstrap resamples, refitting the models with the original penalties in each resample. Predictive performance and grouped Shapley contributions for 2025 and 2026 are reported in Table S1.

### S3.2 Collaborator identification and regression models

We identified collaborators and corresponding authors from authors' first-authored publications in 2023–2024. A collaborator was measurable if at least three first-authored publications remained after excluding publications shared with the focal author and an early resistance score could be calculated from the remaining publications. Collaborators' resistance was calculated using the same method described above. Among initially resistant authors with at least one measurable collaborator of the relevant type, the samples contained 9,649 focal authors for the analysis of all collaborators and 1,642 for the analysis of corresponding-author collaborators. $C_i$ indicated whether an author had at least one measurable collaborator with non-positive early resistance.

For each year and collaborator category, we fitted:

$$R_{it} = \gamma_{0t} + \gamma_{1t} C_i + \gamma_{2t} R_{i,E} + \boldsymbol{Z}_i^T \boldsymbol{\theta}_t + \epsilon_{it} \quad (11)$$

The control vector $\mathbf{Z}_i$ included field, the institutional country category, academic age, historical mean $\alpha$, cumulative and recent output, early publication count, mean team size, and both total and measurable collaborator counts. The coefficient $\gamma_{1t}$ estimates the adjusted difference in subsequent resistance between authors with and without a measurable non-resistant collaborator. Table S2 reports the regression coefficients for both years and collaborator categories.

### S3.3 Comparisons by institutional country category

We compared authors whose affiliation countries during 2023–2024 were all classified as English-speaking or all classified as non-English-speaking, excluding authors with affiliations in both categories. We used all authors in each category as the denominator for both the initial and continuous resistance proportions. Among initially resistant authors, we used ordinary least squares models to compare subsequent resistance, controlling for early resistance, field, academic age, publication history, historical mean α and collaboration characteristics. These models included 32,468 authors from fields with at least 20 initially resistant authors in each category. By 2026, continuously resistant authors accounted for 25.4% of the English-speaking category and 15.4% of the non-English-speaking category (Figure S2a). The regression comparisons also showed lower subsequent resistance in the non-English-speaking category, by 0.027 in 2025 and 0.039 in 2026 (Figure S2b). Using the continuous institutional country category measure and retaining authors with affiliations in both categories yielded the same direction of association.

## Supplementary Note 4: Reference-use measures and comparisons

### S4.1 Reference age and disciplinary composition

Reference age was the citing publication year minus the cited publication year. We computed reference-age measures only when valid publication years were available for at least ten linked references and for at least half of the linked reference list. Valid years ranged from 1000 to the citing year. Reference age composition was measured as the shares of references aged 0–7 years, 8–19 years and at least 20 years. These bands were selected after examining the finer-grained age distribution.

For disciplinary composition, cited publications were assigned to the subfield associated with their primary topic. Let $K_p$ denote the number of represented subfields and $q_{pk}$ the share of classifiable references assigned to subfield k. We calculated:

$$\begin{aligned} V_p &= K_p, \\ B_p &= -\frac{\sum_{k=1}^{K_p} q_{pk} \log q_{pk}}{\log K_p}, \\ D_p &= 1 - \sum_{k=1}^{K_p} q_{pk}^2 . \end{aligned} \tag{12}$$

Variety counts represented subfields, balance measures normalised Shannon evenness, and diversity is the Gini–Simpson index. We computed these measures only when subfields were available for at least ten linked references and for at least half of the linked reference list. Balance was treated as missing when only one subfield was represented.

### S4.2 Reference reuse

Let $d_p$ denote the publication date of paper $p$, $\mathcal{R}_{ip}$ its linked reference set, and $\mathcal{H}_i(d_p)$ the union of references in author $i$'s first-authored publications dated before $d_p$. Reference reuse was calculated as:

$$U_{ip} = \frac{\left|\mathcal{R}_{ip} \cap \mathcal{H}_i(d_p)\right|}{\left|\mathcal{R}_{ip}\right|} \tag{13}$$

Reference reuse required at least ten linked references and was treated as missing when no earlier first-authored publications were observed. Publications on the same date did not establish prior use.

We constructed two fixed reference libraries. The initial five-year library contained references cited in an author's first-authored publications during the five calendar years beginning with the year of the first observed first-authored publication. The five-year window had to end before 2023. The pre-2023 library contained references cited in all first-authored publications before 2023. For each library, we distinguished references in the library, references previously cited outside the library and references first observed in the author's current reference list. The two-group analyses included 7,563 authors for the initial five-year library and 17,847 for the pre-2023 library.

### S4.3 Adjusted comparisons

For each author-level measure $k$ and period $s$, we fitted:

$$Z_{is}^{(k)} = \eta_{0s}^{(k)} + \eta_{1s}^{(k)} G_i + \boldsymbol{C}_i^T \boldsymbol{\theta}_s^{(k)} + \boldsymbol{W}_{is}^{(k)T} \boldsymbol{\lambda}_s^{(k)} + \varepsilon_{is}^{(k)} \tag{14}$$

Here, $G_i = 1$ identifies authors with positive mean resistance in both periods and $G_i = 0$ identifies authors with non-positive mean resistance in both periods. The early and late periods were 2023–2024 and 2025–2026, respectively. Group membership depended on the mean score in each period. The coefficient $\eta_{1s}^{(k)}$ represents the adjusted difference between the groups. Common controls $\mathbf{C}_i$ included characteristics measured during 2023–2024: field, the institutional country category, mean team size, recent output and early publication count. They also included academic age in 2023, historical mean $\alpha$, and publication count and mean reference-list length before 2023. The period-specific controls $\mathbf{W}_{is}^{(k)}$ included publication date and the proportion of each author's publications in each field during the corresponding period. The model for 2025–2026 additionally included the mean institutional country category during 2025–2026. Variety, balance, diversity and reuse models additionally controlled for reference-list length. Balance models also controlled for variety. Reuse models additionally controlled for cumulative and recent output in the corresponding period, measured before each publication. We used heteroskedasticity-robust (HC3) standard errors. Estimates for reference age, age composition and disciplinary composition are reported in Table S3. Estimates for reference reuse and the fixed reference libraries are reported in Table S4.

### S4.4 Comparisons across resistance groups

We extended the reference-use comparisons to include authors whose mean score was positive in 2023–2024 and non-positive in 2025–2026. Figures S3–S5 label this group "Later non-resistant". The "Persistently resistant" and "Consistently non-resistant" groups had positive and non-positive mean scores, respectively, in both periods. For each outcome and period, we fitted an ordinary least squares model with resistance group as a categorical predictor, using the control variables described in S4.3 and giving each author equal weight. We used HC3 standard errors to calculate the 95% confidence intervals for the adjusted means. The reference-use comparisons included 24,177 authors, of whom 10,169 were included in the initial five-year library analysis. The comparisons covered reference age, reference variety, reference balance, reference diversity and reference reuse (Figure S3), reference age composition (Figure S4), and the shares of references belonging to the initial five-year and pre-2023 libraries (Figure S5).

## Supplementary Figures

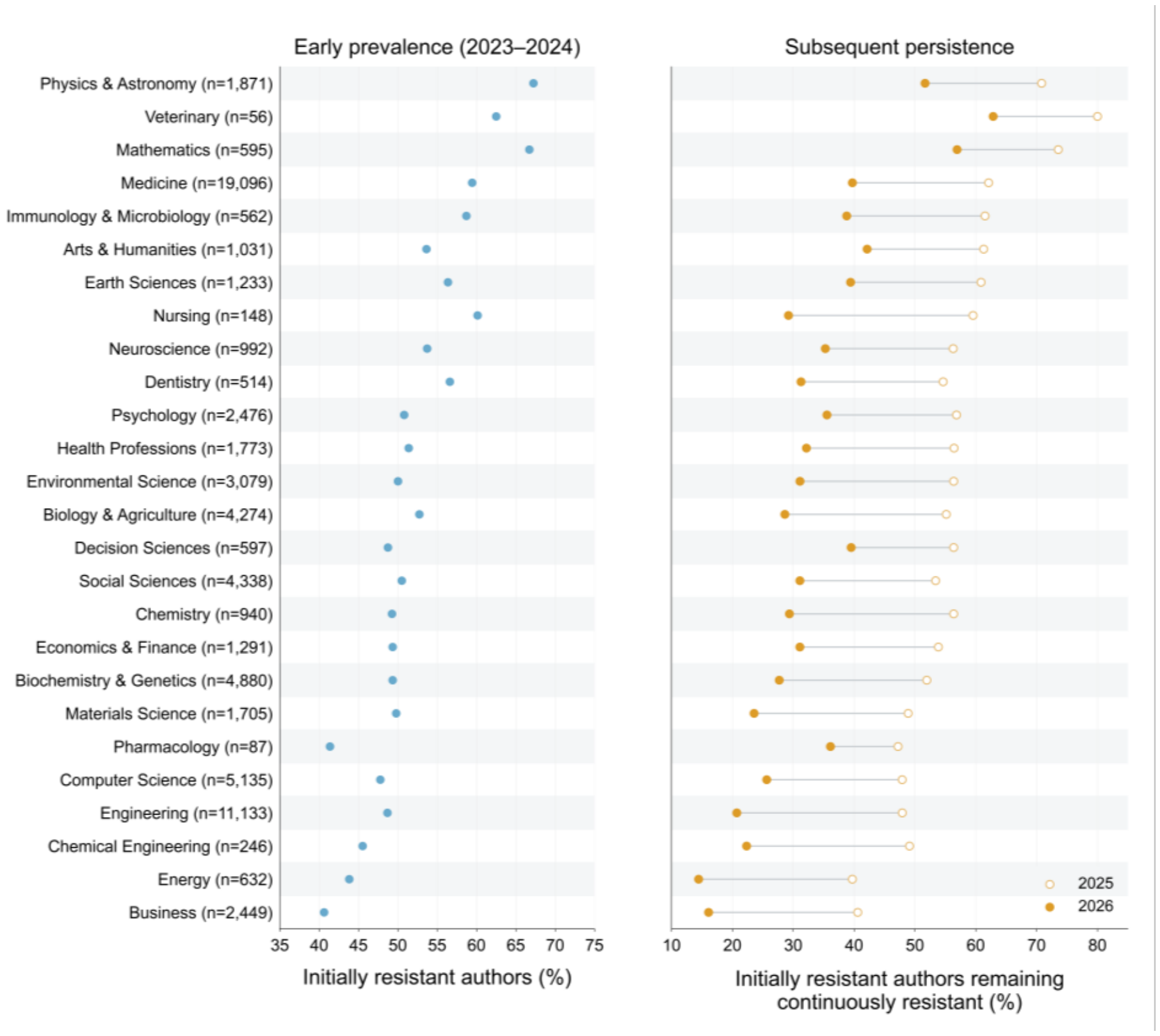

**Figure S1. Disciplinary differences in resistance.** The share of authors with early resistance in each field (left) and the shares of those authors who remained continuously resistant through 2025 and through 2026 (right).

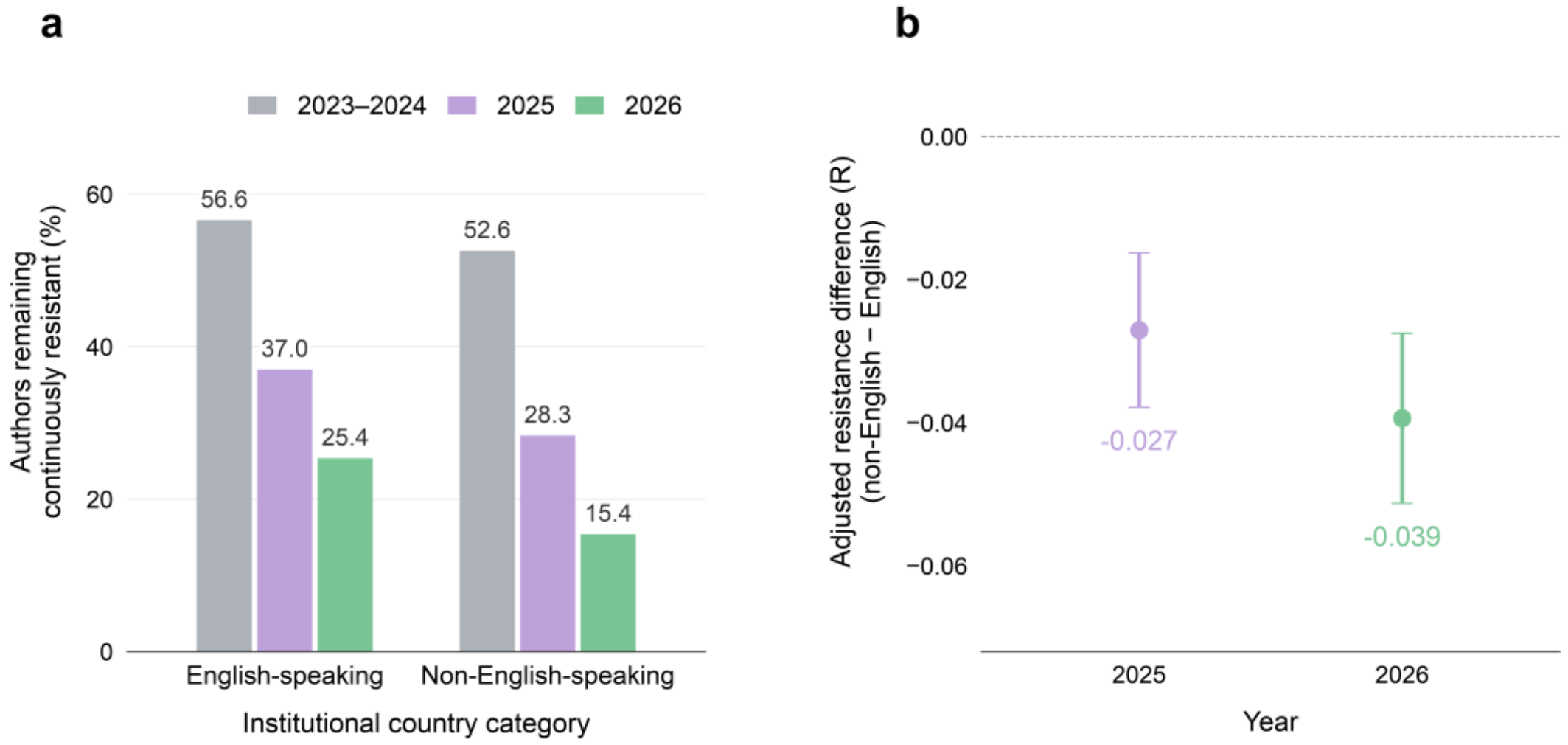

**Figure S2. Resistance by institutional country category. a**, Proportions of all authors in each category with early resistance (2023–2024) or continuous resistance through 2025 or 2026. **b**, Adjusted differences in subsequent resistance between the non-English-speaking and English-speaking categories among initially resistant authors. Error bars indicate 95% confidence intervals.

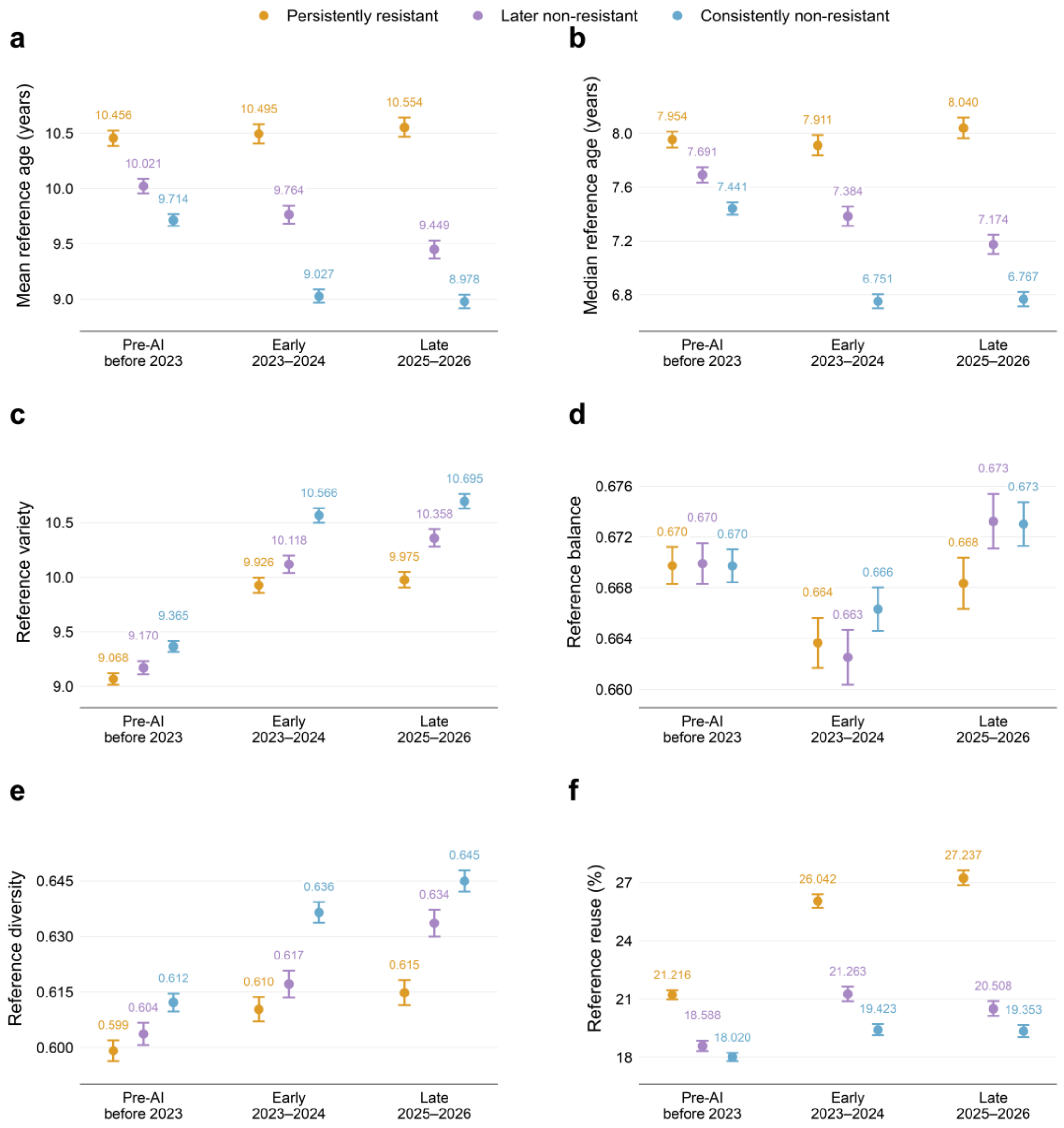

**Figure S3. Reference use across three resistance groups. a**, Mean reference age. **b**, Median reference age. **c**, Reference variety. **d**, Reference balance. **e**, Reference diversity. **f**, Reference reuse. Points show adjusted means, and error bars indicate 95% confidence intervals. Orange indicates positive mean resistance in both periods, purple indicates positive early and non-positive late resistance, and blue indicates non-positive mean resistance in both periods. "Pre-AI" denotes publications before 2023.

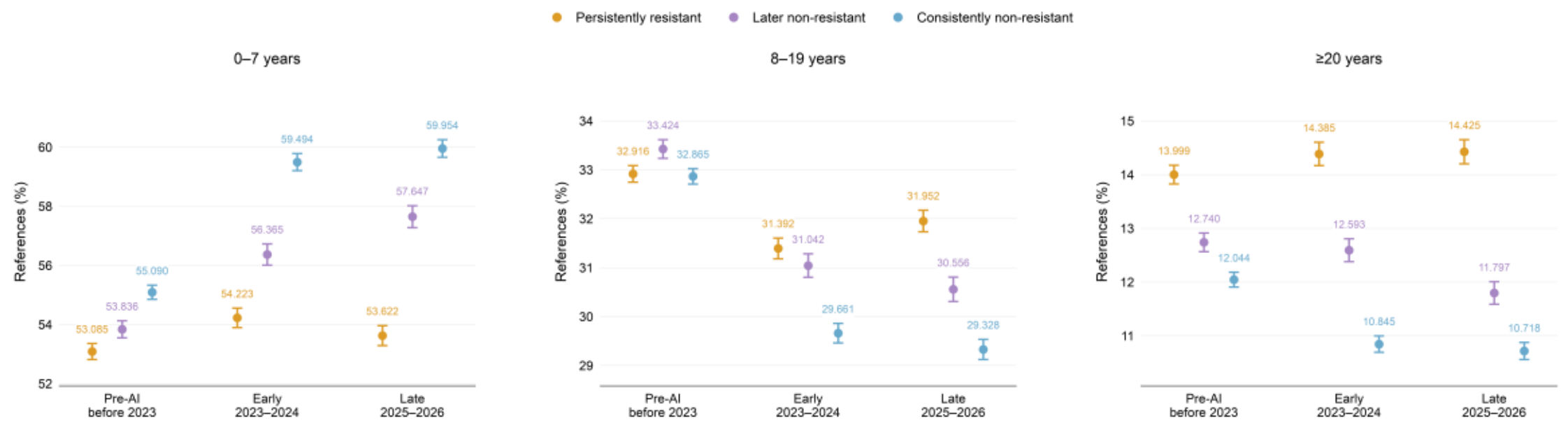

**Figure S4. Reference age composition across three resistance groups.** Panels show the shares of references aged 0–7 years (left), 8–19 years (middle), and at least 20 years (right). Points show adjusted means, and error bars indicate 95% confidence intervals. Orange indicates positive mean resistance in both periods, purple indicates positive early and non-positive late resistance, and blue indicates non-positive mean resistance in both periods. “Pre-AI” denotes publications before 2023.

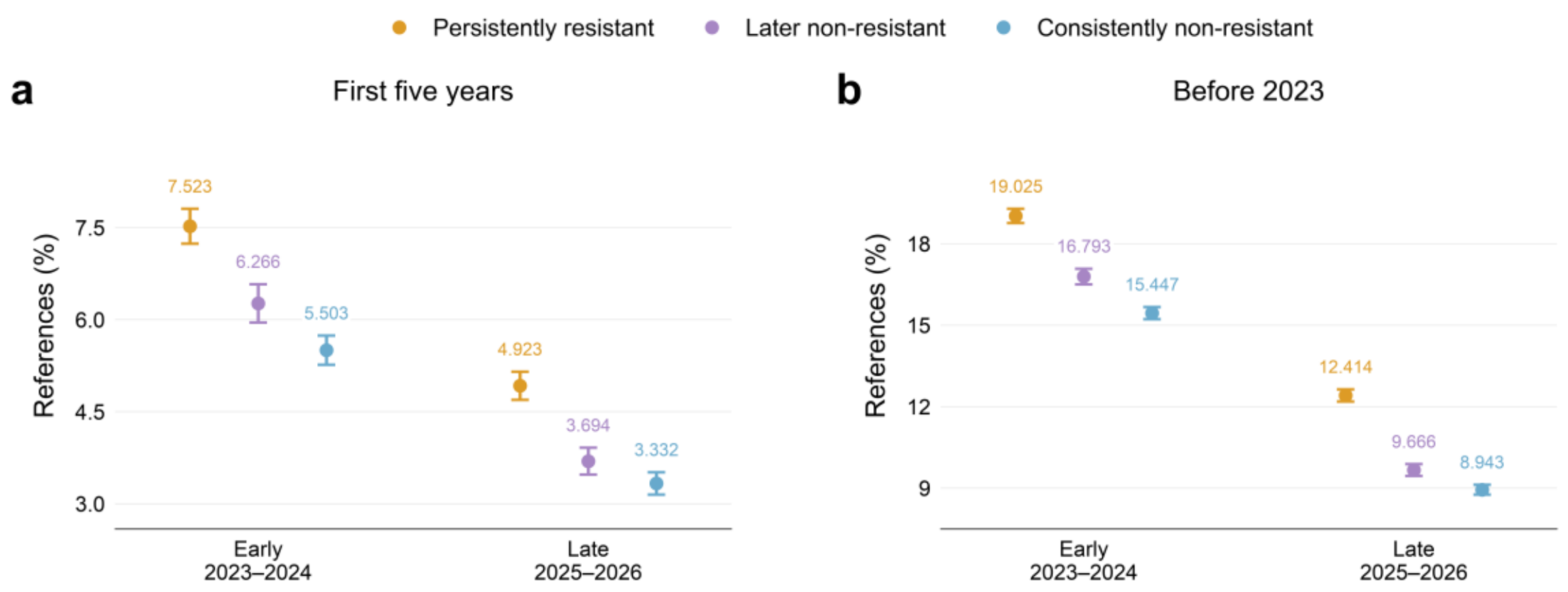

**Figure S5. Use of fixed reference libraries across three resistance groups. a,** Shares of references previously cited in the five calendar years starting with each author's first observed first-authored publication year. **b,** Shares of references cited before 2023. Points show adjusted means, and error bars indicate 95% confidence intervals. Orange indicates positive mean resistance in both periods, purple indicates positive early and non-positive late resistance, and blue indicates non-positive mean resistance in both periods.

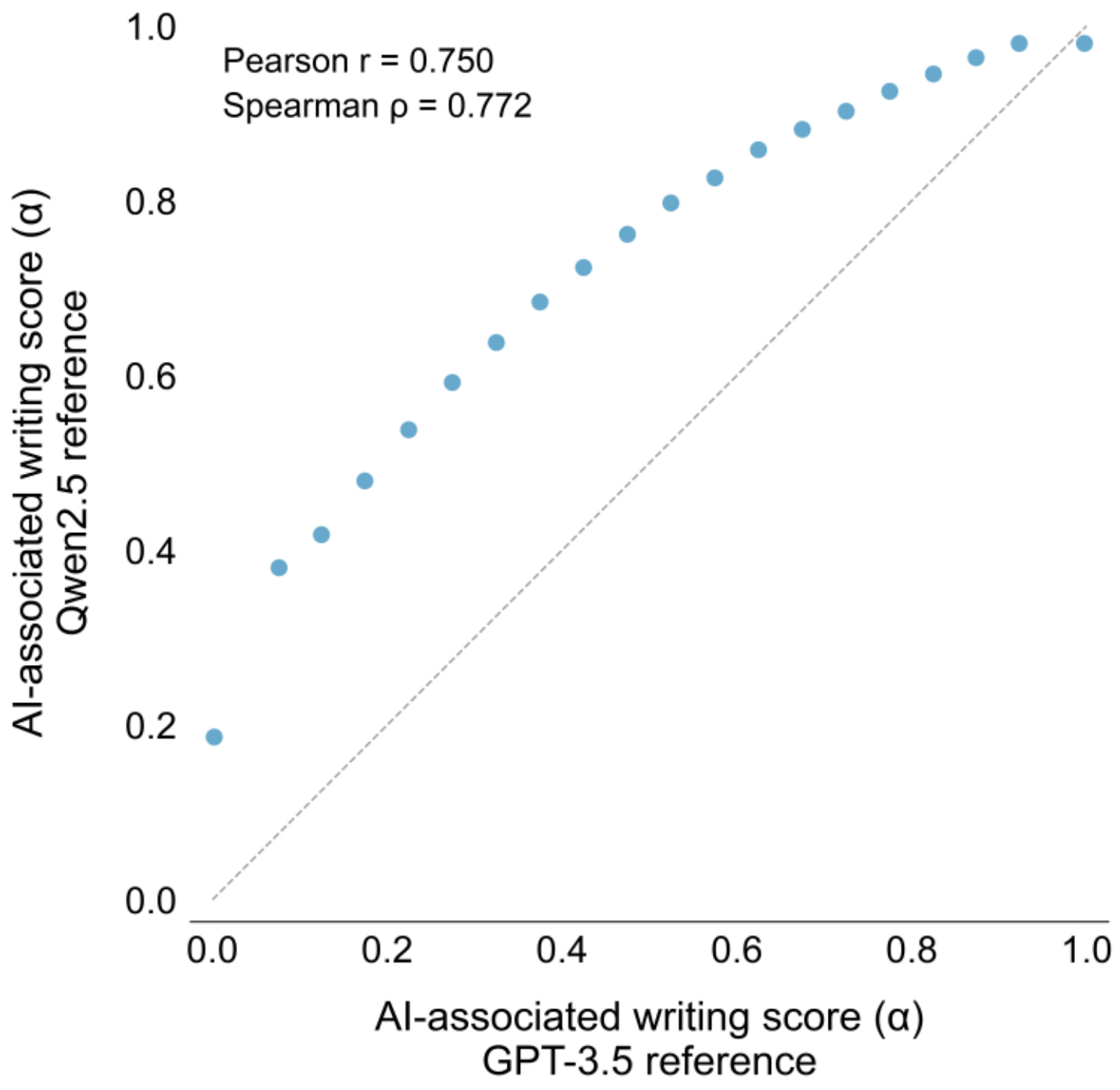

**Figure S6. AI-associated writing scores using alternative reference texts.** Points show mean Qwen-based scores within bins of GPT-3.5-based scores. The diagonal indicates equal scores.

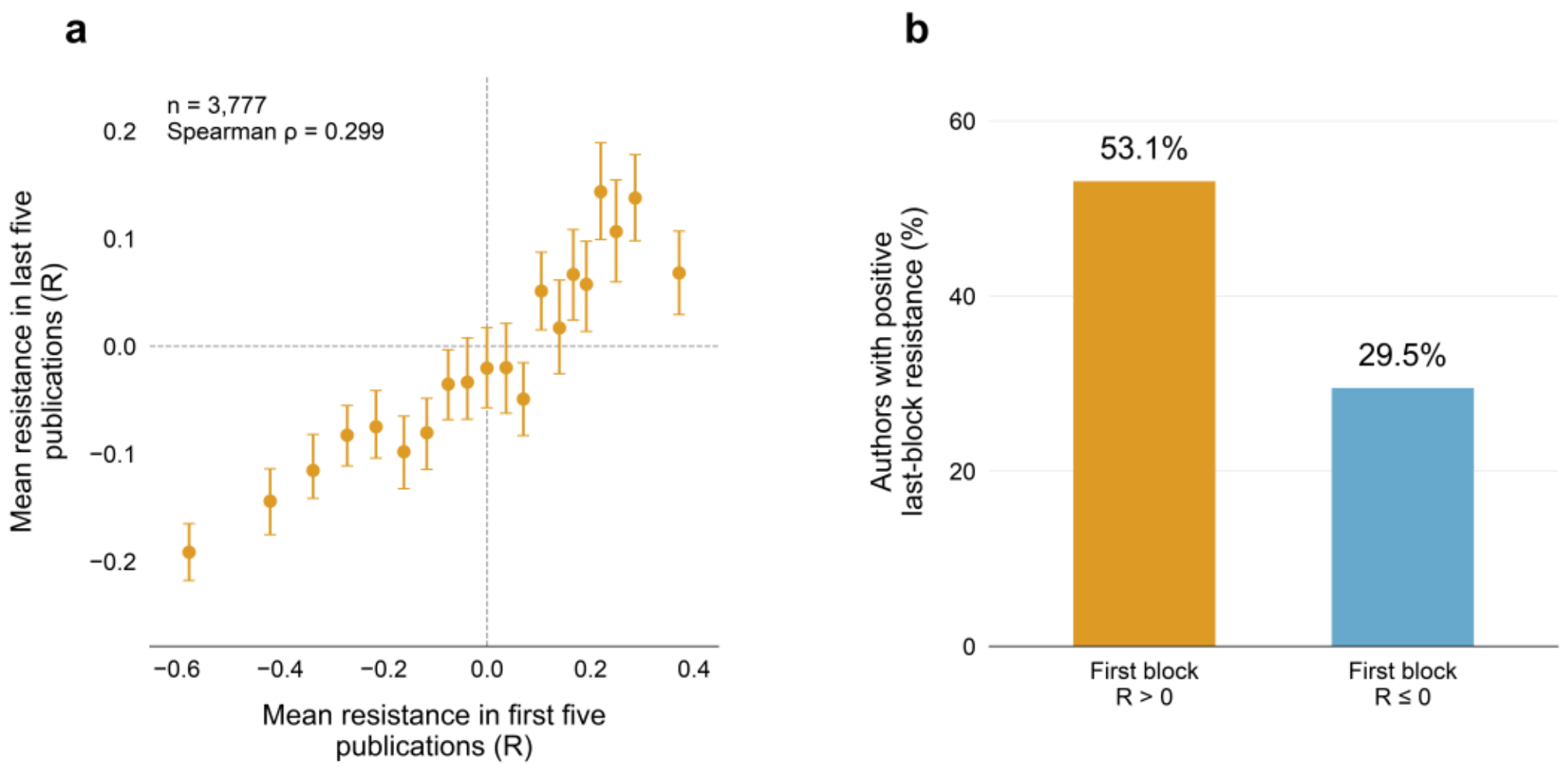

**Figure S7. Resistance across the publication sequence. a**, Binned association between mean resistance in the first and last five publications. Error bars indicate 95% confidence intervals from 500 author-level bootstrap resamples. **b**, Proportions of authors with positive resistance in the last five publications, grouped by whether resistance in the first five was positive or non-positive.

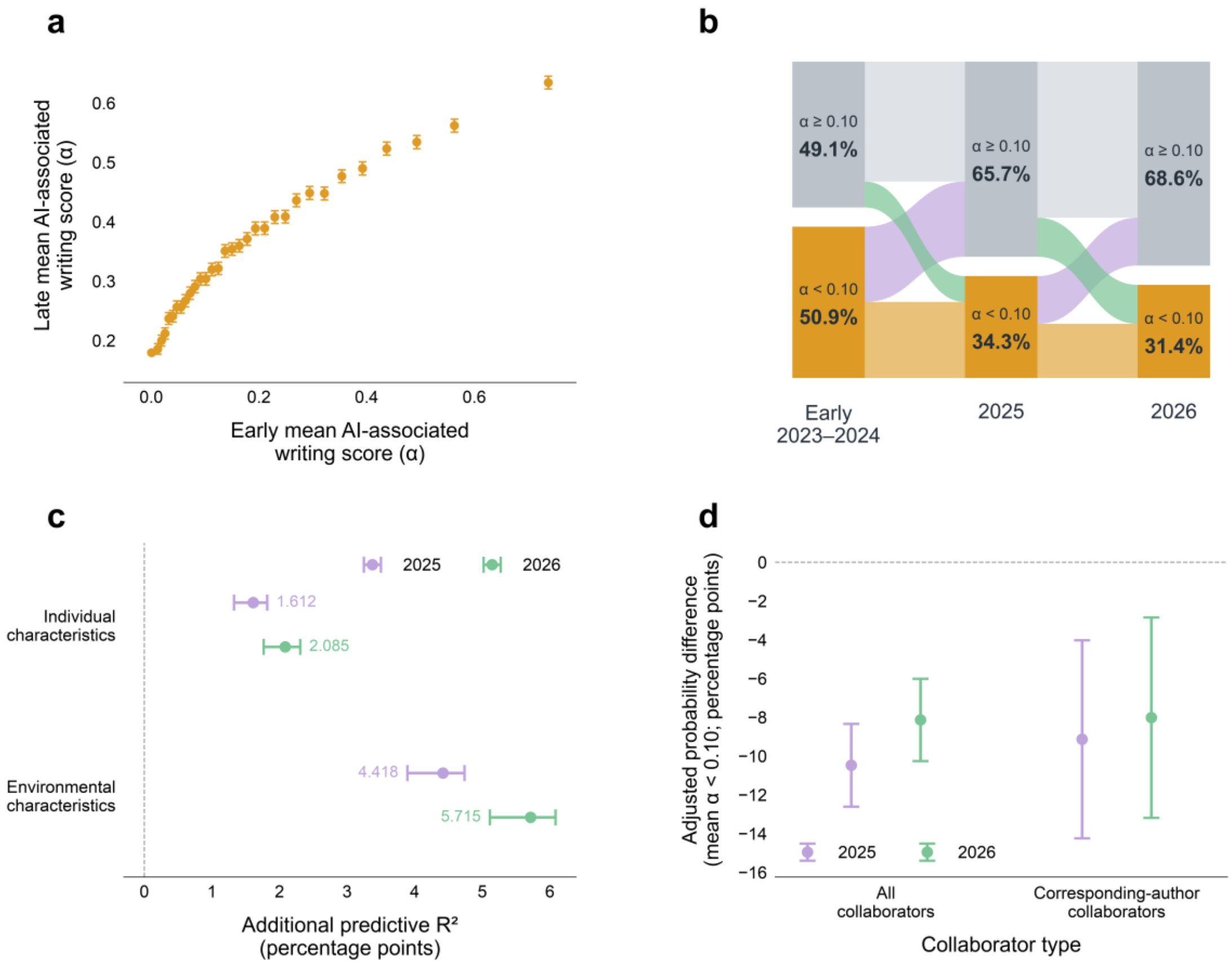

**Figure S8. Analyses using the absolute score criterion.** **a**, Association between early and late AI-associated writing scores. **b,** Transitions between low-score ($\alpha < 0.10$) and higher-score ($\alpha \geq 0.10$) status. The threshold applies to the mean score for 2023–2024 and to each annual mean for 2025 and 2026. Percentages refer to the full cohort at each time point. **c**, Grouped Shapley contributions of individual and environmental characteristics. **d**, Adjusted differences in the probability of subsequent low-score status, by collaborator type.

## Supplementary Tables

**Table S1. Predictive performance and contributions of individual and environmental characteristics. Each analysis included 37,629 authors with positive early resistance. The abbreviation pp denotes percentage points.**

A. Predictive performance

| Year | Model | Out-of-sample $R^2$ [95% CI] | Improvement over baseline, pp [95% CI] |
|---|---|---|---|
| 2025 | Baseline | 0.0240 [0.0207, 0.0269] | — |
| 2025 | Baseline + individual | 0.0269 [0.0231, 0.0299] | 0.295 [0.131, 0.391] |
| 2025 | Baseline + environment | 0.0321 [0.0279, 0.0347] | 0.812 [0.521, 0.957] |
| 2025 | Full model | 0.0400 [0.0347, 0.0426] | 1.604 [1.184, 1.773] |
| 2026 | Baseline | 0.0139 [0.0115, 0.0167] | — |
| 2026 | Baseline + individual | 0.0155 [0.0126, 0.0176] | 0.156 [−0.000, 0.232] |
| 2026 | Baseline + environment | 0.0224 [0.0186, 0.0248] | 0.849 [0.585, 0.983] |
| 2026 | Full model | 0.0271 [0.0226, 0.0294] | 1.319 [0.975, 1.459] |

B. Grouped Shapley contributions

| Year | Individual characteristics, pp [95% CI] | Environmental characteristics, pp [95% CI] | Environment minus individual, pp [95% CI] |
|---|---|---|---|
| 2025 | 0.544 [0.347, 0.671] | 1.061 [0.765, 1.216] | 0.517 [0.262, 0.739] |
| 2026 | 0.313 [0.150, 0.409] | 1.006 [0.738, 1.149] | 0.693 [0.442, 0.891] |

**Table S2. Collaborator associations with subsequent resistance. Values in parentheses are HC3 standard errors. P values were adjusted using the Benjamini–Hochberg procedure across the four tests of collaborator associations.**

| | **All collaborators** | | **Corresponding-author collaborators** | |
|---|---|---|---|---|
| **Variable** | **2025** | **2026** | **2025** | **2026** |
| Non-resistant collaborator | −0.0595 (0.0085) | −0.0515 (0.0092) | −0.0720 (0.0196) | −0.0482 (0.0216) |
| Early resistance | 0.4037 (0.0336) | 0.3725 (0.0356) | 0.4759 (0.0828) | 0.3214 (0.0883) |
| Historical mean $\alpha$ | −0.0299 (0.0553) | −0.0512 (0.0554) | −0.1025 (0.1071) | −0.1021 (0.1139) |
| Institutional country category | 0.0426 (0.0107) | 0.0478 (0.0121) | 0.0305 (0.0302) | 0.0492 (0.0369) |
| Cumulative output | 0.0023 (0.0074) | 0.0067 (0.0080) | −0.0033 (0.0179) | −0.0031 (0.0196) |
| Recent output | −0.0038 (0.0100) | 0.0098 (0.0106) | −0.0383 (0.0250) | 0.0236 (0.0273) |
| Early publication count | 0.1810 (0.0184) | 0.1172 (0.0202) | 0.1392 (0.0375) | 0.0317 (0.0418) |
| Mean team size | 0.1731 (0.0195) | 0.1267 (0.0219) | 0.0369 (0.0248) | 0.0306 (0.0274) |
| Collaborator count | −0.1227 (0.0145) | −0.0970 (0.0161) | −0.0593 (0.0199) | −0.0573 (0.0228) |
| Measurable collaborator count | −0.0264 (0.0152) | −0.0100 (0.0173) | 0.0182 (0.0795) | 0.0483 (0.0841) |
| Field indicators | Yes | Yes | Yes | Yes |
| Academic-age spline | Yes | Yes | Yes | Yes |
| Authors | 9,649 | 9,649 | 1,642 | 1,642 |
| With a non-resistant collaborator | 5,259 | 5,259 | 838 | 838 |
| $R^2$ | 0.0519 | 0.0389 | 0.0683 | 0.0446 |

| | All collaborators | | Corresponding-author collaborators | |
|---|---|---|---|---|
| **Variable** | **2025** | **2026** | **2025** | **2026** |
| 95% CI for collaborator association | [−0.0761, −0.0429] | [−0.0696, −0.0335] | [−0.1105, −0.0336] | [−0.0905, −0.0059] |
| Adjusted *P* for collaborator association | <0.001 | <0.001 | <0.001 | 0.026 |

**Table S3. Adjusted differences in reference age and disciplinary composition. Values are adjusted differences between authors with positive mean resistance in both periods and authors with non-positive mean resistance in both periods, in that order. Brackets contain 95% confidence intervals. The abbreviation pp denotes percentage points.**

A. Reference age and disciplinary composition

| | **Before 2023** | **2023–2024** | **2025–2026** |
|---|---|---|---|
| Mean reference age (years) | 0.749 [0.658, 0.840] | 1.470 [1.363, 1.578] | 1.589 [1.479, 1.699] |
| Median reference age (years) | 0.520 [0.442, 0.599] | 1.163 [1.069, 1.257] | 1.287 [1.190, 1.383] |
| Variety (*V*) | −0.285 [−0.360, −0.211] | −0.612 [−0.711, −0.512] | −0.713 [−0.814, −0.612] |
| Balance (*B*) | −0.0002 [−0.0022, 0.0018] | −0.0030 [−0.0056, −0.0003] | −0.0048 [−0.0076, −0.0021] |
| Diversity (*D*) | −0.0129 [−0.0167, −0.0091] | −0.0254 [−0.0299, −0.0210] | −0.0302 [−0.0347, −0.0256] |

B. Reference age composition

| | **Before 2023** | **2023–2024** | **2025–2026** |
|---|---|---|---|
| References aged 0–7 years (pp) | −2.047 [−2.423, −1.672] | −5.298 [−5.750, −4.845] | −6.393 [−6.862, −5.924] |
| References aged 8–19 years (pp) | 0.060 [−0.181, 0.301] | 1.731 [1.430, 2.031] | 2.639 [2.327, 2.951] |
| References aged ≥20 years (pp) | 1.987 [1.754, 2.220] | 3.567 [3.295, 3.839] | 3.754 [3.470, 4.039] |

**Table S4. Adjusted differences in reference reuse and fixed reference libraries. Values are adjusted differences between authors with positive mean resistance in both periods and authors with non-positive mean resistance in both periods, in that order. Brackets contain 95% confidence intervals. The abbreviation pp denotes percentage points.**

A. Reference reuse

| | **Before 2023** | **2023–2024** | **2025–2026** |
|---|---|---|---|
| Reference reuse (pp) | 3.241<br>[2.909, 3.574] | 6.690<br>[6.215, 7.165] | 7.947<br>[7.426, 8.467] |

B. Initial five-year library

| | **2023–2024** | **2025–2026** |
|---|---|---|
| Fixed-library references (pp) | 2.043<br>[1.663, 2.424] | 1.615<br>[1.309, 1.921] |
| Previously cited outside the fixed library (pp) | 4.782<br>[4.187, 5.376] | 5.995<br>[5.282, 6.709] |
| First-observed references (pp) | −6.825<br>[−7.551, −6.099] | −7.610<br>[−8.421, −6.800] |

C. Pre-2023 library

| | **2023–2024** | **2025–2026** |
|---|---|---|
| Fixed-library references (pp) | 3.629<br>[3.266, 3.992] | 3.429<br>[3.125, 3.733] |
| Previously cited outside the fixed library (pp) | 0.741<br>[0.557, 0.926] | 2.347<br>[2.012, 2.682] |
| First-observed references (pp) | −4.371<br>[−4.783, −3.959] | −5.776<br>[−6.242, −5.310] |